\documentstyle[11pt]{article}

\makeatletter
\def\dif{{\rm d}}
\def\deriv{\@ifnextchar[{\@deriv}{\@deriv[]}}
   \def\@deriv[#1]#2#3{\mathchoice%
{{\dif^{#1}#2\over\dif{#3}^{#1}}}{{\dif^{#1}#2/\dif{#3}^{#1}}}%
{{\dif^{#1}#2\over\dif{#3}^{#1}}}{{\dif^{#1}#2/\dif{#3}^{#1}}}}
\def\derpar#1#2{\mathchoice%
{{\partial#1\over\partial#2}}{{\partial#1/\partial#2}}%
{{\partial#1\over\partial#2}}{{\partial#1/\partial#2}}}
\def\dderpar#1#2#3{\mathchoice%
{{\partial^2 #1\over\partial #2\,\partial #3}}%
{{\partial^2 #1/\partial #2\,\partial #3}}%
{{\partial^2 #1\over\partial #2\,\partial #3}}%
{{\partial^2 #1/\partial #2\,\partial #3}}}
\def\secteqno{\@addtoreset{equation}{section}%
\def\theequation{\thesection.\arabic{equation}}}
\newcounter{subequation}
\def\thesubequation{\alph{subequation}}
\def\sneqnarray{\stepcounter{equation}\let\@currentlabel=\theequation
\setcounter{subequation}{1}
\def\@eqnnum{{\rm (\theequation.\thesubequation)}}
\global\@eqcnt\z@\tabskip\@centering\let\\=\@eqncr\let\@@eqncr=\@@sneqncr
$$\halign to \displaywidth\bgroup\@eqnsel\hskip\@centering
 $\displaystyle\tabskip\z@{##}$&\global\@eqcnt\@ne
 \hskip 2\arraycolsep \hfil${##}$\hfil
 &\global\@eqcnt\tw@ \hskip 2\arraycolsep $\displaystyle\tabskip\z@{##}$\hfil
  \tabskip\@centering&\llap{##}\tabskip\z@\cr}
\def\endsneqnarray{\@@sneqncr\egroup $$\global\@ignoretrue}
\def\@@sneqncr{\let\@tempa\relax
   \ifcase\@eqcnt \def\@tempa{& & &}\or \def\@tempa{& &}
   \else \def\@tempa{&}\fi
     \@tempa \if@eqnsw\@eqnnum\stepcounter{subequation}\fi
     \global\@eqnswtrue\global\@eqcnt\z@\cr}
\def\nobiblabels{\def\@lbibitem[##1]##2{\@bibitem{##2}}}
\makeatother
\def\tabaddress#1{{\small\it\begin{tabular}[t]{c}#1\\[1.2ex]\end{tabular}}}
\def\artit#1{``#1'',}
\def\ben{\begin{enumerate}}
\def\een{\end{enumerate}}
\def\beq{\begin{equation}}
\def\eeq{\end{equation}}
\def\bea{\begin{eqnarray}}
\def\eea{\end{eqnarray}}
\def\beann{\begin{eqnarray*}}
\def\eeann{\end{eqnarray*}}
\def\beasn{\begin{sneqnarray}}
\def\eeasn{\end{sneqnarray}}

\newtheorem{teor}{Theorem}
\newtheorem{prop}{Proposition}

\def\UBECM{Departament d'Estructura i Constituents de la Mat\`eria\\
   Universitat de Barcelona\\
   Av.~Diagonal 647\\
   08028 Barcelona\\
   Catalonia, Spain}
\def\UPCMAT{Departament de Matem\`atica Aplicada i Telem\`atica\\
   Universitat Polit\`ecnica de Catalunya\\
   Campus Nord, edifici C3\\
   08071 Barcelona\\
   Catalonia, Spain}

\def\Real{{\bf R}}
\def\Ker{\mathop{\rm Ker}\nolimits}
\def\Tan{{\rm T}}

\def\temp{\hbox{\rm D}_t}

\def\coorsup#1#2#3{{#1}^{#2}, \ldots, {#1}^{#3}}
\def\coorinf#1#2#3{{#1}_{#2}, \ldots, {#1}_{#3}}

\def\map#1{\mathrel{\mathop{\to}\limits^{#1}}}
\def\feble#1{\mathrel{\mathop{\simeq}\limits_{#1}}}
\def\forta#1{\mathrel{\mathop{\cong}\limits_{#1}}}

\def\FL{{\cal F}\!L}
\def\pot{{\cal F}}

\def\GH{G_{\rm H}}
\def\GI{G_{\rm I}}
\def\GL{G_{\rm L}}
\def\deltaH{\delta_{\rm H}}
\def\PFC{\hbox{\sc pfc}}
\def\KE{K_E}

\def\bm#1{\hbox{\boldmath$#1$}}
\def\bfe{\bm e}
\def\p{\bm p}
\def\hatp{\hat{\bm p}}
\def\x{\bm x}

\def\nor#1{(#1#1)}
\def\delt#1{\Delta_{#1}}
\def\ort#1{\bfe_{#1}}

\def\P#1#2{P_{#1}^{(#2)}}
\def\lligam#1#2{\phi_{#1}^{#2}}

\let\eps=\varepsilon
\title{Gauge transformations for higher-order lagrangians}
\author{\sc
Xavier Gr\`acia$^{\sharp}$
and
Josep M. Pons$^\flat$%
\\
\tabaddress{$^\sharp$\UPCMAT}
\\
\tabaddress{$^\flat$\UBECM}
\\[2mm]
\small e-mails: {\tt
xgracia@mat.upc.es\quad
pons@rita.ecm.ub.es}
}
\date{August 1995}

\def\ppclassif{{\noindent\tt
hep-th/9509094\\
UB-ECM-PF 95/15\\}
PACS: 03.20.+i  \qquad  Mathematics s.c.: 70H}

\nobiblabels
\secteqno

\begin{document}
\thispagestyle{empty}
\maketitle

\begin{abstract}
Noether's symmetry transformations for higher-order
lagrangians are studied.
A characterization of these transformations is presented,
which is useful to find gauge transformations
for higher-order singular lagrangians.
The case of second-order lagrangians is studied in detail.
Some examples that illustrate our results are given;
in particular, for the lagrangian of a relativistic particle
with curvature, lagrangian gauge transformations are obtained,
though there are no hamiltonian gauge generators for them.
\end{abstract}

\vfill
\ppclassif
\clearpage

\section{Introduction}

Among the symmetries of a classical dynamical system described through
an action principle, Noether's symmetries
\cite{Bes-Noether,Hil-Noether,Noe-Noether,Olv-DiffEq}
(i.e., those that leave the action invariant, up to boundary terms)
play a central role.
They are the usual symmetries considered in systems of physical interest,
their characterization is very simple,
and, most importantly, they are the kind of symmetries that
we must consider when dealing with quantum systems;
this is clear from the path
integral formulation,
where the main ingredient is the classical action together with
the measure in the space of field configurations.

Here we will consider continuous symmetries, either rigid or gauge.
In the latter case, the infinitesimal transformation will depend upon
arbitrary functions of time ---in mechanics--- or space-time ---in field
theory.  In order for these gauge transformations to exist
the lagrangian must be singular.
In a first-order lagrangian this means that
the hessian matrix with respect to the velocities is singular;
it is so with respect to the highest derivatives in a higher-order case.
Constants of motion appear associated to rigid symmetries whereas
first-class hamiltonian constraints appear associated to gauge symmetries
\cite{Dir-lectures};
in this case Noether's identities also appear.

For regular lagrangians the constant of motion associated with a
Noether's symmetry is in fact the generator of the symmetry when
expressed in hamiltonian formalism.
For singular lagrangians this statement is not always true:
a lagrangian Noether's transformation may not be projectable
to phase space.

In
\cite{BGGP-Noether}
and
\cite{GP-Noether}
several aspects of Noether's symmetries for first-order lagrangians
have been studied;
in particular the projectability of these transformations from lagrangian
to hamiltonian formalism.
Let us explain this point.
Let $L(q,\dot q)$ be a first-order lagrangian and
$\FL$ its associated Legendre transformation mapping velocity space
to phase space:
$\FL(q, \dot q) = (q, \hat p)$,
where $\hat p(q,\dot q) = \derpar{L}{\dot q}$ are the momenta.
Given a Noether's symmetry
$\delta q(t,q,\dot q)$ of~$L$,
the corresponding constant of motion $\GL(t,q,\dot q)$
turns out to be projectable to a function $\GH(t,q,p)$ in phase space.
This means that there is a function $\GH$ whose pull-back $\FL^*(\GH)$
through the Legendre transformation is $\GL$;
in other words,
$\GH(t,q,\hat p) = \GL(t,q,\dot q)$.
(Notice that, for a singular lagrangian, not every function
in velocity space is projectable to a function in phase space,
due to the singularity of the Legendre transformation.)
Then there is a simple characterization of the functions $\GH$ that
correspond to a Noether's symmetry
\cite{GP-Noether}.
Finally, the function
$\GH$ acts as a kind of generator for the Noether's symmetry $\delta q$.
If the functions $\delta q(t,q,\dot q)$ are projectable to phase space,
then $\GH$ can be chosen
(between the functions whose pull-back to velocity space is~$\GL$)
such that it generates the symmetry in the same way as for
regular lagrangians, i.e., through Poisson bracket;
otherwise, $\GH$ still generates the Noether's symmetry
though not in such a simple way.

In this paper we extend these results to higher-order lagrangians.
For these lagrangians there exists a hamiltonian formulation,
due to Ostrogradski\u i;
in the case of singular lagrangians, Dirac's theory may be applied,
and for instance the search for generators of symmetry transformations
\cite{GP-gauge}
is performed as for the first-order case.
As we will see,
when we look for Noether's symmetry transformations
of a higher-order singular lagrangian
the situation is rather different from the first-order case.
The most remarkable difference is that in the higher-order case
the constant of motion $\GL$ is not necessarily projectable
to a function $\GH$ in phase space.

To perform this analysis we make use of the results of
\cite{BGPR-second}
and
\cite{GPR-higher}.
As it will be summarized in section~2,
given a $k$-th order lagrangian
there are $k-1$ intermediate spaces
$P_0 \map{\pot_0} P_1 \map{} \ldots P_{k-1} \map{\pot_{k-1}}
P_k$
between those of lagrangian ($P_0$) and hamiltonian ($P_k$)
formalisms,
where $\pot_0$, \ldots, $\pot_{k-1}$ are the
``partial Legendre-Ostrogradski\u{\i}'s transformations''.
So the study of the projectability of a lagrangian quantity (in $P_0$)
to phase space ($P_k$) is more involved.
In particular, unlike the first-order case,
the constant of motion of a Noether's symmetry,
although being projectable to the intermediate space $P_1$,
is not necessarily projectable to the phase space.

Our characterization of Noether's transformations
is especially relevant when looking for gauge transformations.
For instance, in
\cite{GP-Noether}
there is a lagrangian not possessing hamiltonian gauge generators,
but such that our method provides lagrangian gauge symmetries for it.
Another example is given by the lagrangian of a conformal particle
\cite{GR-conformal}:
it has a hamiltonian gauge symmetry that can not be written
in a covariant form despite the covariance of the hamiltonian constraints;
in this case our method allows to construct
a covariant lagrangian gauge symmetry.
In this paper a similar behaviour is shown to occur
in a second-order lagrangian,
namely the curvature of the world-line of a relativistic particle:
it will be shown that it has no hamiltonian gauge generators,
but two independent lagrangian gauge transformations
will be obtained for it.

The paper is organized as follows.
In section~2 some results on higher-order lagrangians
are summarized.
In section~3 Noether's transformations for higher-order lagrangians
are studied,
and a characterization of them is introduced.
In section~4 the case of second-order lagrangians
is developed in full detail.
As an application of these results,
in section~5 the example of the particle with the curvature as a
lagrangian is studied;
other examples are also studied in the next section.
The paper ends with a section with conclusions and
an appendix about hamiltonian symmetry transformations.

\section{Higher-order lagrangians}

Here we present some results and notation from reference
\cite{GPR-higher}.
See also
\cite{BGPR-second,CSC-higher,LR-higher,Sau-jets}
for higher-order lagrangians and
higher-order tangent bundles.

Let $Q$ be an $n$-dimensional differentiable manifold
with coordinates%
\footnote{Indices of coordinates are usually suppressed.}
$q = q_0$.
On its higher-order tangent bundles
$\Tan^rQ$
we consider natural coordinates
$(\coorinf q0r)$.
A $k$th order lagrangian is a function $L\colon \Tan^kQ \to \Real$.

The Ostrogradski\u{\i}'s momenta are
\beq
\hat p^i =
\sum_{j=0}^{k-i-1} (-1)^j \temp^j \left( \derpar{L}{q_{i+j+1}} \right) ,
\qquad 0\le i\le k-1 ,
\eeq
where $\temp = \derpar{}{t} + \sum_{i} q_{i+1} \derpar{}{q_i}$
is the total time-derivative.
Equivalently,
\beq
{\hat p}^{k-1} = \derpar{L}{q_k} , \qquad
{\hat p}^{i-1} = \derpar{L}{q_i} - \temp {\hat p}^i .
\label{ppunt}
\eeq
Notice that $\hat p^i$ depends only on $\coorinf q 0 {2k-1-i}$.

In coordinates the Euler-Lagrange equations can be written
$[L]_{q(t)} = 0$, with
\bea
[L] &=&
\sum_{r=0}^{k} (-1)^r \temp^r \left( \derpar{L}{q_r} \right)
\nonumber
\\
&=&
\derpar{L}{q_0} - \temp {\hat p}^0
\label{EL-p}
\\
&=&
\alpha - (-1)^{k-1} q_{2k} W ,
\label{EL-W}
\eea
where
$$
\alpha = \derpar{L}{q_0}
- q_1 \derpar{{\hat p}^0}{q_0} - \ldots
- q_{2k-1} \derpar{{\hat p}^0}{q_{2k-2}}
$$
and $W$ is the hessian matrix with respect to the highest-order
velocities,
$$
W = \dderpar{L}{q_k}{q_k} .
$$

Introducing the momenta step-by-step,
for $0 \le r \le k$ an
intermediate space $P_r$
can be defined, with coordinates
$ (\coorinf q0{2k-1-r}; \coorsup p0{r-1}) $.
In particular,
the lagrangian and hamiltonian spaces are
$P_0 = \Tan^{2k-1}Q$ and
$P_k = \Tan^*(\Tan^{k-1}Q)$.
Observe that $P_k$ has a canonical Poisson bracket,
for which $\{q_r^i,p^s_j\} = \delta_r^s \delta^i_j$.

The partial Ostrogradski\u{\i}'s transformations
$ \pot_r\colon P_r \to P_{r+1} $
can be introduced,
with local expression
\beq
\pot_r(\coorinf q0{2k-1-r}; \coorsup p0{r-1}) =
        (\coorinf q0{2k-2-r}; \coorsup p0{r-1}, \hat p^r).
\eeq
The ``total'' Legendre-Ostrogradski\u{\i}'s transformation is
$\FL = \pot_{k-1} \circ \ldots \circ \pot_0
\colon P_0 \to P_k$.

On $P_r$ there exists an unambiguous evolution operator
$K_r$,
which is a vector field along $\pot_r$,
$K_r \colon P_r \to \Tan(P_{r+1})$,
satisfying certain conditions
\cite[theorem 4]{GPR-higher}.
In coordinates it reads
\begin{eqnarray}
K_r &=&
q_1 \derpar{}{q_0} + \ldots + q_{2k-1-r} \derpar{}{q_{2k-2-r}} +
\nonumber \\
 & & \null +
\left( \derpar{L}{q_0} \right) \derpar{}{p^0} +
\left( \derpar{L}{q_1} - p^0 \right) \derpar{}{p^1} +
\ldots +
\left( \derpar{L}{q_r} - p^{r-1} \right) \derpar{}{p^r}.
\nonumber \\
 & &
\label{Kc}
\end{eqnarray}
The various evolution operators are connected by
\beq
K_{r-1} \cdot \pot_r^*(g) = \pot_{r-1}^*(K_r \cdot g) ,
\label{K-alpha}
\eeq
for $1 \leq r \leq k-1$;
here $\pot_r^*(g)$ denotes the pull-back of $g$ through~$\pot_r$.

These intermediate evolution operators
act as differential operators from functions in~$P_{r+1}$
to functions in~$P_r$.
They can be extended to act on time-dependent functions;
for instance, given a time-dependent function in $P_1$,
$g(t,\coorinf q 0 {2k-2},p^0)$,
$$
K_0 \cdot g =
\pot_0^*\left(\derpar gt\right) + q_1 \pot_0^*\left(\derpar g{q_0}\right)
+ \ldots +
q_{2k-1} \pot_0^*\left(\derpar g{q_{2k-2}}\right) +
\derpar{L}{q_0} \pot_0^*\left(\derpar g{p^0}\right) .
$$
By computing $K_0 \cdot g - \temp \pot_0^*(g)$
using the chain rule
an interesting relation is obtained:
\beq
K_0 \cdot g = [L]\pot_0^*\left(\derpar g{p^0}\right) + \temp \pot_0^*(g) .
\label{Ktemp}
\eeq

We assume that $W$ has constant rank $n-m$.
Then the $\pot_r$ have constant rank
$2kn-m$, since
$$
\derpar{\hat p^r}{q_{2k-1-r}} = (-1)^{k-1-r} W ,
$$
and $\P{r+1}1 := \pot_r(P_r)$
is then assumed to be a closed submanifold of $P_{r+1}$
locally defined by $m$ independent primary constraints
$\lligam{r+1}{\mu}$.
The primary hamiltonian constraints
---those defining $\P k1$---
can be chosen to be independent of $\coorsup p0{k-2}$.
Then the primary constraints of $P_r$
can be obtained by applying $K_r$ to the primary constraints
of $P_{r+1}$
\cite[proposition 9]{GPR-higher}:
\beq
\lligam{r}{\mu} := K_r \cdot \lligam{r+1}{\mu} .
\label{K-lligam}
\eeq
This is also true for $r=0$.
Indeed one can write evolution equations on each space $P_r$
($0 \leq r \leq k-1$);
these equations are equivalent to the Euler-Lagrange equations.
The first consistency conditions for these equations
are just the constraints $\lligam{r}{\mu}$ defined above.

The primary constraints yield a basis for $\Ker W$:
$$
\gamma_\mu =
\pot_{k-1}^*\left( \derpar{\lligam{k}{\mu}}{p^{k-1}} \right) ,
$$
which can also be written as
$
(-1)^{k-r}\pot_{r-1}^*\left( \derpar{\lligam{r}{\mu}}{p^{r-1}} \right)
$,
provided that the $\lligam{r}{\mu}$ are defined by (\ref{K-lligam}).
Notice that $\gamma_\mu$ depends only on $(\coorinf q0k)$.
Then, a basis for $\Ker \Tan(\pot_r)$ is constituted by the vector fiels
$$
\Gamma^r_\mu = \gamma_\mu \derpar{}{q_{2k-1-r}} .
$$
These can be used to test the projectability of a function in $P_r$
to $P_{r+1}$:
$\Gamma^r_\mu \cdot g = 0$.

We notice also the commutation relations
$$
\Gamma^r_\mu \cdot(K_r \cdot g) = \pot_r^*(\Gamma^{r+1}_\mu \cdot g) ,
$$
for $0 \leq r \leq k-1$,
where $\Gamma^k_\mu$ is understood as
$\Gamma^k_\mu \cdot g = \{g, \lligam{k}{\mu}\}$.

Using the null vectors $\gamma_\mu$, (\ref{Ktemp}) and (\ref{K-lligam}),
we obtain in particular the primary lagrangian constraints as
$$
\lligam{0}{\mu} = K_0 \cdot \lligam{1}{\mu} =
(-1)^{k-1} [L]\gamma_\mu =
(-1)^{k-1} \alpha\gamma_\mu .
$$

There is a hamiltonian function in $P_k$,
which is a projection of the lagrangian energy function
$E_0 (q_0, \ldots, q_{2k-1}) =
\hat p^0 q_1 + \ldots
+ \hat p^{k-1} q_k - L(\coorinf q0k)$;
it can be chosen in the particular form
\beq
H = \sum_{r=0}^{k-2} p^r q_{r+1} + h(\coorinf q0{k-1}; p^{k-1}) .
\label{hamiltoniana-c}
\eeq
The usual presymplectic (Dirac's) analysis can be performed in $\P k1$.
In fact,
there are stabilization algorithms for the dynamics of the
intermediate spaces
and all the constraints in $P_r$
---not only the primary ones---
are obtained applying $K_r$ to all the constraints in~$P_{r+1}$
\cite[theorem 8]{GPR-higher}.
This result holds indeed at each step of the stabilization algorithms.

\section{Noether's transformations}

An infinitesimal Noether's symmetry
\cite{Bes-Noether,Hil-Noether,Noe-Noether,Olv-DiffEq}
(see also
\cite{CF-timedep,CLM-convNoether,CLM-higherNoether,%
FP-Noether,Li-higher,Lus-Noether})
is an infinitesimal transformation $\delta q$ such that
$$
\delta L = \temp F ,
$$
for a certain~$F$.
It yields a conserved quantity
$G = \sum_{r=0}^{k-1} \hat p^r \delta q_r - F$,
where $\delta q_r = \temp^r \delta q$,
since
$$
[L]\delta q + \temp G = 0 ;
$$
this is proved using the Euler-Lagrange equations (\ref{EL-p})
and the relation between the momenta.

So let us consider a $\delta q(t,\coorinf q0{2k-1})$,
and a function $\GL(t,\coorinf q 0 {2k-1})$ such that
\beq
[L]\delta q + \temp \GL = 0 .
\label{noetherG}
\eeq
Notice that the highest derivative in this relation,
$q_{2k}$, appears linearly, and its coefficient is
$$
(-1)^k W \delta q - \derpar \GL{q_{2k-1}} = 0 ;
$$
so, contracting with the null vectors $\gamma_\mu$ we obtain that
$$
\Gamma_\mu \cdot \GL = 0 ,
$$
that is to say, $\GL$ is projectable to a function $\GI$ in~$P_1$,
$$
\GL = \pot_0^*(\GI) .
$$
Now, using (\ref{Ktemp}), (\ref{noetherG}) becomes
$$
[L] \left( \delta q - \pot_0^*\left(\derpar \GI{p^0}\right) \right)
+ K_0 \cdot \GI = 0 .
$$
Looking again at the coefficient of $q_{2k}$ in this expression,
we obtain
$$
W \left( \delta q - \pot_0^*\left(\derpar \GI{p^0}\right) \right) = 0 ,
$$
and so the parentheses enclose a null vector of~$W$:
$$
\delta q - \pot_0^*\left(\derpar \GI{p^0}\right) = \sum_\mu r^\mu
\gamma_\mu $$
for some $r^\mu(t,\coorinf q0{2k-1})$.
Substituting this expression we obtain
\beq
K_0 \cdot \GI + \sum_\mu r^\mu (\alpha\gamma_\mu) = 0 .
\label{noetherK}
\eeq

So we have proved the following result:

\begin{teor}
Let $\delta q(t,q_0,\ldots,q_{2k-1})$
be a Noether's transformation with conserved quantity~$\GL$.
Then $\GL$ is projectable to a function $\GI$ in $P_1$ such
that\footnote{%
$f \feble{M} 0$ means $f=0$ on~$M$ (Dirac's weak equality).}
\beq
K_0 \cdot \GI \feble{\P01} 0 ,
\eeq
where $\P01$ is the primary lagrangian constraint submanifold.

Conversely, given a function $\GI(t,q_0,\ldots,q_{2k-2},p^0)$
satisfying this relation,
if $r^\mu$ are functions such that
$K_0 \cdot \GI = - \sum_\mu r^\mu (\alpha\gamma_\mu)$
then
\beq
\delta q = \pot_0^*\left(\derpar{\GI}{p^0}\right) + \sum_\mu r^\mu
\gamma_\mu \label{deltaq}
\eeq
is a Noether's transformation with conserved quantity
$\GL = \pot_0^*(\GI)$.
\end{teor}

Notice that $\delta q$ is not necessarily projectable
to $P_1$, not to mention to phase space $P_k$;
in fact, the projectability of $\delta q$
is equivalent to the projectability of the functions~$r^\mu$.

There is also a certain indetermination in the functions
$r^\mu$
\cite{GP-nocons}.
For instance, if there are at least two primary lagrangian constraints
then one can add convenient combinations of these constraints
to the $r^\mu$,
namely, an antisymmetric combination of the primary lagrangian constraints,
in a way that (\ref{noetherK}) is still satisfied;
however, this change corresponds to adding a trivial gauge transformation
\cite{Hen-afBRST}
to the original transformation,
and so we still have the same transformation on-shell
(i.e., for solutions of the equations of motion).
Another interesting case occurs when the primary lagrangian
constraints are not independent;
in
\cite{GP-nocons}
the relation between this fact and Noether's transformations
with vanishing conserved quantity is studied.
For instance, one of the primary lagrangian constraints,
say $\chi = K_0 \cdot \psi$,
may be identically vanishing,
and so for $\GI=0$ {\it any}\/ value for the corresponding $r$
is admissible to fulfill $K_0 \cdot \GI + r\chi = 0$.
This yields a Noether's transformation
$$
\delta q = r \,\pot_0^*\left( \derpar{\psi}{p^0} \right) ;
$$
for instance $r$ might be an arbitrary function of time,
thus yielding a gauge transformation.
Summing up:
unlike the case of a regular lagrangian,
where there is a one-to-one correspondence between Noether's
transformations and conserved quantities,
for a singular lagrangian in general there is a whole family
of Noether's transformations associated with a single conserved quantity.

\section{Projectability of Noether's transformations in the case
of second-order lagrangians}

In the first-order case, $k=1$, the results of the previous section
tell us that $\GL$ is projectable to the phase space $P_1 = \Tan^*Q$.
As we will see shortly this is not true for a higher-order case
$k \geq 2$.
This means that there is no guarantee that we can write the
conserved quantity in canonical variables, let alone to get the
Noether's transformation in phase space:
as we can read off from (\ref{deltaq}),
this is not always possible even for the first-order case.

In order to clarify both issues, projectability of $\GL$ and
projectability of $\delta q$, which in fact we will see that are
related, we will perform a thorough study of the case $k=2$, which
will already show the basic features of the general picture for any~$k$.

Let us consider from now on the case $k=2$.
A basis for
$\Ker \Tan(\FL)$ is given
\cite{BGPR-second}
by the vector fields
\beann
\Gamma^0_{\mu_1} &=&
\gamma_{\mu_1} \derpar{}{q_{3}} ,
\\
\tilde\Gamma^0_{\mu_1'} &=&
\gamma_{\mu_1'}\derpar{}{q_{2}} + \eta_{\mu_1'}\derpar{}{q_{3}} .
\eeann
The index $\mu_1'$ is a part of
the indices $\mu_1$ that corresponds to the splitting of
the primary hamiltonian constraints $\lligam{2}{\mu_1}$ into
the first class ones, $\lligam{2}{\mu_1'}$,
and the second class ones $\lligam{2}{\mu_1''}$.
The function $\eta_{\mu_1'}$ can be written as
$\eta_{\mu_1'} = \derpar{\lligam{2}{\mu_2}}{p^1}$,
where $\lligam{2}{\mu_2} = \{\lligam{2}{\mu_1'},H\}$
are the secondary constraints in phase space
(here $\mu_1'$ and $\mu_2$ run over the same set of indices, but are
distinguished in order to label primary or secondary constraints)

It is easy to prove that the vector fields
$\tilde\Gamma^0_{\mu_1}$ are projectable to the intermediate space $P_1$.
In fact, since the
definition of $\Ker\Tan(\FL)$ requires that
$\tilde\Gamma^0_{\mu_1'} (\pot_0^*(p^0)) = 0$, we get immediately
$\tilde\Gamma^0_{\mu_1'} \circ \pot_0^*  =
\pot_0^* \circ \Gamma^1_{\mu_1'}$
(as operators on functions of the intermediate space).

Now we can check the condition of projectability of $\GL$ to $P_2$.
Since $\Gamma^0_{\mu_1} \cdot \GL = 0$, we only have to check whether
$\tilde\Gamma^0_{\mu_1'} \cdot \GL$ vanishes:
\bea
\tilde\Gamma^0_{\mu_1'} \cdot \GL &=&
\tilde\Gamma^0_{\mu_1'} \cdot \pot_0^*(\GI) =
\pot_0^*(\Gamma^1_{\mu_1'} \cdot \GI) =
\Gamma^0_{\mu_1'} \cdot (K_0 \cdot \GI) =
\nonumber
\\
&=&
\Gamma^0_{\mu_1'} \cdot (-r^\mu(\alpha \gamma_\mu)) =
-(\Gamma^0_{\mu_1'} \cdot r^\mu)(\alpha \gamma_\mu) =
\nonumber
\\
&=&
-\alpha ((\Gamma^0_{\mu_1'} \cdot r^\mu)\gamma_\mu) =
-\alpha (\Gamma^0_{\mu_1'} \cdot (r^\mu \gamma_\mu)) =
-\alpha (\Gamma^0_{\mu_1'} \cdot \delta q) .
\label{projGL}
\eea
Notice that the projectability of $\GL$ to~$P_1$ depends on~$\delta q$.
In this argument we have used several commutation properties
of the $\Gamma$'s, but there are two details to point out.

First,
$\Gamma^0_{\mu_1'} \cdot (\alpha \gamma_\mu) = 0$;
this is a consequence of a more general result:
$$
-\Gamma^0_{\nu} \cdot (\alpha \gamma_\mu) =
\FL^*\{ \lligam{2}{\mu},\lligam{2}{\nu} \} ,
$$
whose proof is immediate:
$$
-\Gamma^0_{\nu} \cdot (\alpha \gamma_\mu) =
\Gamma^0_{\nu} \cdot (K_0 \cdot \lligam{1}{\nu}) =
\pot_0^*(\Gamma^1_{\nu} \cdot (K_1 \cdot \lligam{2}{\nu})) =
\pot_0^*(\pot_1^*(\Gamma^2_{\nu} \cdot \lligam{2}{\nu})) =
\pot_0^*(\pot_1^*\{ \lligam{2}{\mu},\lligam{2}{\nu} \}) .
$$
In particular, taking one of the constraints to be first-class,
$\{ \lligam{2}{\mu},\lligam{2}{\nu'} \} \feble{\P{2}{1}} 0$,
the result is zero.

Second,
$(\Gamma^0_{\mu_1'} \cdot r^\mu)\gamma_\mu =
\Gamma^0_{\mu_1'} \cdot (r^\mu \gamma_\mu)$,
which is trivially true since the vector functions $\gamma_\mu$
are projectable.

Therefore we have obtained an expression for
$\tilde\Gamma^0_{\mu_1'} \cdot \GL$,
and in general it can be different from zero.
Notice
that a sufficient condition for the projectability of $\GL$ to $P_2$ is
that $\delta q$ be projectable to $P_1$.
Notice also that the quantity
$\alpha (\Gamma^0_{\mu_1'} \cdot \delta q)$
is insensitive to the
indetermination of the functions $r^\mu$ which is mentioned at the end
of the previous section.

Now we are going to consider that the conditions are met for the
projectability of $\GL$ to a function $\GH$ in $P_2$,
$\FL^*(\GH) = \GL$.
The function $\GH$ has a certain degree of arbitrariness
because we can add to
it arbitrary combinations of the primary as well as the secondary
constraints in $P_2$.
Let us extract consequences from our assumption.
The function $\pot_1^*(\GH)$ is one of the possible functions $\GI$
considered in the previous section and
therefore we can apply to it the results already obtained there.
In particular:
$$
K_0 \cdot(\pot_1^*(\GH)) \feble{\P01} 0 .
$$
But since $K_0 \circ \pot_1^* = \pot_0^* \circ K_1$,
$$
\pot_0^* (K_1 \cdot \GH) \feble{\P01} 0,
$$
which means that
$$
K_1 \cdot \GH =
\sum_{\mu_1} u_1^{\mu_1} \lligam{1}{\mu_1} +
\sum_{\mu_2} v_1^{\mu_2} \lligam{1}{\mu_2} .
$$
Here
$\lligam{1}{\mu_1}$ and $\lligam{1}{\mu_2}$
are respectively the primary and the secondary constraints is $P_1$
(remenber that $\mu_2$ runs over the same indices as $\mu_1'$).
Notice that
$\pot_0^*(\lligam{1}{\mu_1}) = 0$ and
$\pot_0^*(\lligam{1}{\mu_2}) = (\alpha \gamma_{\mu_1'})$.
Therefore:
$$
K_0 \cdot\pot_1^*(\GH) = \pot_0^*(K_1 \cdot \GH) =
\pot_0^*(v_1^{\mu_1'}) (\alpha\gamma_{\mu_1'}) ,
$$
and, according to the results of the previous section, the
transformation
\beq
\delta q =
\pot_0^*\left( \derpar{\pot_1^*(\GH)}{p^0} \right) +
\sum_{\mu_1'} \pot_0^*(v_1^{\mu_1'}) \gamma_{\mu_1'}
\label{deltaqi}
\eeq
is a Noether's transformation which is projectable to $P_1$.

If we define
$$
\GI = \pot_1^*(\GH) - \sum v_1^{\mu_1'} \lligam{1}{\mu_1'} ,
$$
since $\derpar{\lligam{1}{\mu_1'}}{p^0}
= \gamma_{\mu_1'}$, then
$\delta q = \pot_0^*\left( \derpar{\GI}{p^0} \right)$ and
$$
K_0 \cdot \GI = 0 ,
$$
where we have used
$K_0 \lligam{1}{\mu_1'} = -(\alpha\gamma_{\mu_1'})$.

\begin{prop}
Let $\GL$ be the conserved quantity of a Noether's transformation.
The following statements are equivalent:
\ben
\item
$\GL$ is projectable to a function $\GH$ in $P_2$.
\item
$\GL$ is projectable to a function $\GI$ in $P_1$ such that
$K_0 \cdot \GI = 0$
(and then
$\displaystyle \delta q = \pot_0^*\left( \derpar{\GI}{p^0}\right)$
is a Noether's transformation with conserved quantity~$\GL$.)
\item
Among the family of Noether's transformations whose conserved quantity
is~$\GL$,
there is one transformation $\delta q$ which is projectable to~$P_1$.
\een
\end{prop}

The proof of the equivalence between the first and the second items
is a direct consequence of the discussion preceding the proposition.
Their equivalence to the third item follows also immediately
from (\ref{projGL}).

\smallskip

Now let us consider the case when $\delta q$ is not only projectable to
$P_1$ but also to $P_2$.
This means that $v_1^{\mu_2}$
in (\ref{deltaqi})
is projectable to $P_2$,
$v_1^{\mu_2} = \pot_1^*(v_2^{\mu_2})$.
In such a case, taking into account that
$K_1 \cdot \lligam{2}{\mu_2} = \lligam{1}{\mu_2}$,
the function
$\GH' := \GH - \sum v_2^{\mu_2} \lligam{2}{\mu_2}$
satisfies
\beq
K_1  \cdot \GH' \feble{\P{1}{1}} 0 ,
\label{k1g2}
\eeq
and $\delta q$ can be expressed as
$$
\delta q = \FL^* \left( \derpar{\GH'}{p^0} \right) ,
$$
which explicitly shows the projectability of $\delta q$ to $P_2$.

There is still another way to write (\ref{k1g2}).
If we define
$\KE = \pot_0^* \circ K_1 = K_0 \circ \pot_1^*$,
then (\ref{k1g2}) can be rewritten as
$$
\KE \cdot \GH' = 0.
$$
The definition of $\KE$ allows to rewrite it as
$$
\KE \cdot g =
[L]\,\FL^*\left(\derpar{g}{p^0}\right) + \temp\,\FL^*(g) .
$$
This makes obvious in a direct way that
$\delta q = \FL^* \left( \derpar{\GH'}{p^0} \right)$
is a Noether's transformation.

At this point we have the following result:

\begin{prop}
Let $\GL$ be the conserved quantity of a Noether's transformation.
The following statements are equivalent:
\ben
\item
$\GL$ is projectable to a function $\GH$ in $P_2$ such that
$K_E \cdot \GH = 0$
(or equivalently
$K_1 \cdot \GH \feble{\P{1}{1}} 0$)
(and then
$\displaystyle \delta q = \FL^*\left( \derpar{\GH}{p^0}\right)$
is a Noether's transformation with conserved quantity~$\GL$.)
\item
Among the family of Noether's transformations whose conserved quantity
is~$\GL$,
there is one transformation $\delta q$ which is projectable to~$P_2$.
\een
\end{prop}

Now there is a subtle point.
Is there a hamiltonian symmetry $\deltaH$ such that
$\displaystyle \deltaH q = \derpar{\GH}{p^0}$?
As it is explained in the appendix,
this is true only when (\ref{kg}) is also satisfied,
and so we have the following result:

\begin{prop}
Let $\GH$ be a function in~$P_2$.
The following statements are equivalent:
\ben
\item
$K_1 \cdot \GH = 0$.
\item
$\GH$ is the generator of a hamiltonian symmetry transformation
such that $\delta q = \FL^*(\deltaH q)$,
where $\deltaH q = \{q_0,H\}$,
is a Noether's transformation with conserved quantity~$\FL^*(\GH)$.
\een
\end{prop}

This result can be directly generalized to any lagrangian of order
$k \geq 2$:
the condition for a function $\GH$ in $P_k$
to be a generator of a Noether's hamiltonian symmetry is
\beq
K_{k-1} \cdot \GH = 0 .
\eeq

To summarize this section,
we have started with a general lagrangian Noether's transformation
and we have examined some conditions to be satisfied by it,
each one more restrictive,
the latter being that of a Noether's hamiltonian symmetry
transformation.
Therefore a conserved quantity of a Noether's transformation
lays in one of the four different cases
depicted by the previous propositions.

\section{Application to the particle with curvature}

Given a path $\x(t)$ in Minkowski space~$\Real^d$,
we write $\x_n$ for its $n$th time-derivative,
and $\ort n$ for the vectors obtained by orthogonalizing
---if possible--- the vectors $\x_1$, $\x_2$, \ldots\
For instance,
\beasn
\ort1 &=& \x_1 ,
\\
\ort2 &=& \x_2 - \frac{(\x_2\ort1)}{\nor{\ort1}} \ort1 ,
\\
\ort3 &=& \x_3 - \frac{(\x_3\ort2)}{\nor{\ort2}} \ort2 -
\frac{(\x_3\ort1)}{\nor{\ort1}} \ort1 .
\eeasn
We also write $\delt n$ for the Gramm determinant of the vectors
$\x_1 \ldots \x_n$:
$$
\delt n = \det((\x_i\x_j))_{1 \leq i,j \leq n} .
$$

For a relativistic particle
we consider a lagrangian proportional to the curvature
of its world line
\cite{BGPR-second,Nes,Pis,Plyu},
\beq
L = \alpha \frac{\sqrt{\delt2}}{\delt1}
= \alpha \frac{\sqrt{(\x_1\x_1)(\x_2\x_2)-(\x_1\x_2)^2}}{(\x_1\x_1)} ,
\eeq
where $\alpha$ is a constant parameter.

Obviously $\ort1$, $\ort2$, $\ort3$ are mutually orthogonal.
Moreover,
$$
(\ort2 \x_2) = \nor{\ort2} = \frac{\delt2}{\delt1} , \qquad
(\ort2 \x_3) = \frac{\dot \delt2}{2\delt2} ,
$$
$$
(\ort3 \x_3) = \nor{\ort3} = \frac{\delt3}{\delt2} .
$$
We shall also need
$$
\dot{\ort2} = \ort3 +
\left( \frac{\dot\delt2}{2\delt2}-\frac{\dot\delt1}{2\delt1} \right)
  \ort2 -
\frac{\delt2}{\delt1^2} \ort1 .
$$

The partial Ostrogradski\u\i's transformations are
$$
\begin{array}{rcccl}
P_0 = \Tan^3(\Real^d)    & \stackrel{\pot_0}{\longrightarrow} &
P_1                      & \stackrel{\pot_1}{\longrightarrow} &
P_2 = \Tan^*(\Tan(\Real^d))
\\
(\x_0,\x_1,\x_2,\x_3)    & \mapsto &
(\x_0,\x_1,\x_2,\hatp^0) &         &
\\
                         &         &
(\x_0,\x_1,\x_2,\p^0)    & \mapsto &
(\x_0,\x_1,\p^0,\hatp^1) ,
\end{array}
$$
where the momenta are defined by
\bea
\hatp^1 &:=&
\derpar{L}{\x_2} =
\frac{\alpha}{\sqrt{\delt2}} \ort2 ,
\\
\hatp^0 &:=&
\derpar{L}{\x_1} - \temp \hatp^1 =
-\frac{\alpha}{\sqrt{\delt2}} \ort3 ;
\eea
for the last computation we have used
$$
\derpar{L}{\x_1} =
- \frac{\alpha}{\sqrt{\delt2}} \left(
    \frac{\dot\delt1}{2\delt1} \ort2 + \frac{\delt2}{\delt1^2} \ort1
  \right) .
$$

More precisely, $P_0$ is not all $\Tan^3(\Real^d)$,
but the open subset defined by $\delt1>0$, $\delt2>0$.
Then the vectors $\x_1$ and $\x_2$ are linearly independent,
and so are $\ort1$ and $\ort2$.
Similar remarks hold for $P_1$ and~$P_2$.

The singularity of the partial Ostrogradski\u\i's transformations
is due to the singularity of the hessian matrix
$$
W := \dderpar{L}{\x_2}{\x_2} =
- \derpar{\hatp^0}{\x_3} = \derpar{\hatp^1}{\x_2} .
$$
In our case,
\beq
W_{\mu\nu} = \frac{\alpha}{\sqrt{\delt2}} \left(
\eta_{\mu\nu} -
\frac{{\ort1}_\mu}{\sqrt{\nor{\ort1}}} \frac{{\ort1}_\nu}{\sqrt{\nor{\ort1}}} -
\frac{{\ort2}_\mu}{\sqrt{\nor{\ort2}}} \frac{{\ort2}_\nu}{\sqrt{\nor{\ort2}}}
\right) ,
\eeq
whose rank is $d-2$ in its domain.

The intermediate evolution operators are
\bea
K_1 &:=&
\x_1 \derpar{}{\x_0} +
\x_2 \derpar{}{\x_1} +
\derpar{L}{\x_0} \derpar{}{\p^0} +
\left( \derpar{L}{\x_1} - \p^0 \right) \derpar{}{\p^1}
\nonumber
\\
&=&
\x_1 \derpar{}{\x_0} +
\x_2 \derpar{}{\x_1} -
\left( \p^0 +
  \frac{\alpha}{\sqrt{\delt2}}  \left(
    \frac{\delt2}{\delt1^2}\ort1 + \frac{\dot\delt1}{2\delt1}\ort2
  \right)
\right) \derpar{}{\p^1} ,
\\
K_0 &:=&
\x_1 \derpar{}{\x_0} +
\x_2 \derpar{}{\x_1} +
\x_3 \derpar{}{\x_2} +
\derpar{L}{\x_0} \derpar{}{\p^0}
\nonumber
\\
&=&
\x_1 \derpar{}{\x_0} +
\x_2 \derpar{}{\x_1} +
\x_3 \derpar{}{\x_2} .
\eea

And the Euler-Lagrange equations (in $P_0$) are
$$
[L] = \derpar{L}{\x_0} - \temp \hatp^0 = 0 .
$$

\subsection{Constraints}

The energy in $P_1$ is
$E_1 := (\p^0\x_1) + (\hatp^1\x_2) - L(\x_0,\x_1,\x_2) = (\p^0\x_1)$,
so we take as a hamiltonian
\beq
H = (\p^0\x_1) .
\eeq

Due to the rank of the hessian matrix~$W$,
the definition of $\hatp^1$
---the last partial Ostrogradski\u\i's transformation---
introduces two constraints in the hamiltonian space~$P_2$.
These constraints are obtained immediately from the relations
satisfied by~$\ort2$, and we take them as
\beasn
\phi_2^1 &=&
(\p^1\x_1) ,
\\
\psi_2^1 &=&
\frac12 \left( (\p^1\p^1) - \frac{\alpha^2}{\nor{\x_1}} \right) .
\eeasn
We have
$$
\{\phi_2^1,\psi_2^2\} = 2 \psi_2^1 .
$$

Proceeding with the hamiltonian stabilization we obtain
secondary constraints
\beasn
\phi_2^2 &=& \{\phi_2^1,H\} = -(\p^0\x_1) = -H ,
\\
\psi_2^2 &=& \{\psi_2^1,H\} = -(\p^0\p^1) ,
\eeasn
for which
$$
\left(
\begin{array}{cc}
\{\phi_2^1,\phi_2^2\} & \{\phi_2^1,\psi_2^2\} \\
\{\psi_2^1,\phi_2^2\} & \{\psi_2^1,\psi_2^2\}
\end{array}
\right)
=
\left(
\begin{array}{cc}
-\phi_2^2 & \psi_2^2 \\
-\psi_2^2 & \frac{\alpha^2}{\nor{\x_1}^2} \phi_2^2
\end{array}
\right) .
$$

Finally we obtain a tertiary constraint
\beq
\psi_2^3 = \{\psi_2^2,H\} = (\p^0\p^0) ,
\eeq
whereas $\{\phi_2^2,H\} = \{\psi_2^3,H\} = 0$.
The Poisson bracket of $\psi_2^3$ with the primary constraints
is zero.

Notice that all the constraints are first class,
but the Poisson bracket between the two secondary constraints
is the tertiary constraint:
$$
\{\phi_2^2,\psi_2^2\} = \psi_2^3 .
$$

\medskip

The constraints in $P_1$ are obtained by applying the operator
$K_1$ to the hamiltonian constraints.
We have:
\beann
K_1 \cdot \phi_2^1 &=:&
\phi_1^1 = -(\p^0 \x_1) ,
\\
K_1 \cdot \psi_2^1 &=:&
\psi_1^1 = -(\p^0 \hatp^1) =
 -\frac{\alpha}{\sqrt{\delt2}} (\p^0 \ort2) ,
\\
K_1 \cdot \phi_2^2 &=&
\frac{(\x_1 \x_2)}{\nor{\x_1}} \phi_1^1 +
\frac{\sqrt{\delt2}}{\alpha} \psi_1^1
\feble{} 0 ,
\\
K_1 \cdot \psi_2^2 &=&
(\p^0 \p^0)
- \frac{(\x_1 \x_2)}{\nor{\x_1}} \psi_1^1
- \frac{\alpha\sqrt{\delt2}}{\nor{\x_1}^2} \phi_1^1
\feble{} (\p^0 \p^0) =: \psi_1^2 ,
\\
K_1 \cdot \psi_2^3 &=&
0
\eeann
Instead of defining $\psi_1^2 = K_1 \cdot \psi_2^2$
we prefer, for simplicity, to use $\psi_1^2 = (\p^0 \p^0)$,
which defines the same constraint submanifold.
With this convention,
$\pot_1^*(\phi_2^{i+1}) = \phi_1^{i}$,
$\pot_1^*(\psi_2^{i+1}) = \psi_1^{i}$.

\medskip

Similarly from the intermediate constraints
$\phi_1^1$, $\psi_1^1$ and $\psi_1^2$
we obtain the lagrangian constraints:
\beann
K_0 \cdot \phi_1^1 &=& 0 ,
\\
K_0 \cdot \psi_1^1 &=:&
\psi_0^1 = (\hatp^0 \hatp^0) ,
\\
K_0 \cdot \psi_1^2 &=& 0 .
\eeann
Again
$\pot_0^*(\phi_1^{i+1}) = \phi_0^{i}$,
$\pot_0^*(\psi_1^{i+1}) = \psi_0^{i}$.

\medskip

{}From the expression of the hessian matrix
it is obvious that $\Ker W = \langle \ort1, \ort2 \rangle$.
Indeed,
\beann
\gamma_\phi &=& \pot_1^* \left( \derpar{\phi_2^1}{\p^1} \right) =
 -\pot_0^* \left( \derpar{\phi_1^1}{\p^0} \right) = \x_1 ,
\\
\gamma_\psi &=& \pot_1^* \left( \derpar{\psi_2^1}{\p^1} \right) =
 -\pot_0^* \left( \derpar{\psi_1^1}{\p^0} \right) = \hatp^1 .
\eeann

\subsection{Hamiltonian gauge transformations}

We are going to show that the model does not have
any hamiltonian gauge transformation
constructed from a generating function.

According to the appendix,
we look for a generator of the form (\ref{G}),
and apply the algorithm (\ref{algor}).
We first consider
\beq
G_0 = f \phi^1 + g \psi^1 ,
\eeq
with $f$ and $g$ functions to be determined.

Then
\beq
G_1 = -f \phi^2 - g \psi^2 + f' \phi^1 + g' \psi^1 ,
\eeq
for certain $f'$,~$g'$.
We compute
\beann
\{\phi^1,G_1\} &=&
(f-\{\phi^1,f\}) \phi^2 - (g+\{\phi^1,g\}) \psi^2
+\PFC ,
\eeann
\beann
\{\psi^1,G_1\} &=&
-\left( \frac{\alpha^2}{\nor{\x_1}^2} g + \{\psi^1,f\} \right) \phi^2 +
(f-\{\psi^1,g\}) \psi^2
+\PFC ,
\eeann
and so to fulfill the test (\ref{algor}.c)
the expressions in parentheses must be weakly vanishing.

Now
\beq
G_2 = f \phi^3 + g \psi^3 + (\{f,H\}-f') \phi^2 + (\{g,H\}-g') \psi^2 +
 f'' \phi^1 + g'' \psi^1
\eeq
for some $f''$,~$g''$.
The test for $G_2$ requires to compute
\beann
\{\phi^1,G_2\}
&=&
\{\phi^1,g\} \psi^3    +
\left( \{\phi^1,\{f,H\}-f'\}-\{f,H\}+f' \right) \phi^2 +
\\
&& +
\left( \{\phi^1,\{g,H\}-g'\}+\{g,H\}-g' \right) \psi^2
+\PFC ,
\eeann
\beann
\{\psi^1,G_2\}
&=&
\{\psi^1,g\} \psi^3  +
\left(
  \{\psi^1,\{f,H\}-f'\} + \frac{\alpha^2}{\nor{\x_1}^2}(\{g,H\}-g')
\right) \phi^2 +
\\
&& +
\left( \{\psi^1,\{g,H\}-g'\} - \{f,H\} + f' \right) \psi^2
+\PFC .
\eeann
In order that these expressions be strongly primary first class constraints,
the coefficients of $\psi^3$, $\phi^2$ and $\psi^2$
must be weakly vanishing.
{}From the coefficients of $\psi^3$ we obtain in particular that
$\{\phi^1,g\}$ and $\{\psi^1,g\}$
are weakly vanishing.
Looking at the coefficients of $\psi^2$ in the test for $G_1$
we obtain that $f$ and $g$ are weakly vanishing,
so that the generator $G$ is strongly vanishing:
it becomes ineffective, since it leaves all solutions invariant.

\subsection{Lagrangian gauge transformations}

The model has two independent Noether's gauge transformations.

One of them is just the reparametrization.
It arises easily from the fact that
$\phi_0^1 := K_0 \cdot \phi_1^1 = 0$,
{\it i.e.},
one of the primary lagrangian constraints is identically vanishing.
This fact yields a Noether's transformation
with vanishing conserved quantity, $\GL = 0$.
According to the discussion on these transformations,
we obtain a gauge transformation
\beq
\delta\x = \eps(t) \x_1 ,
\eeq
since $\gamma_\phi=\x_1$;
this is just a reparametrization.

The other transformation comes from
$\GI = \eps(t) \psi_1^2 = \eps(t) (\p^0\p^0)$,
for which
$\GL = \eps(t) \psi_0^1 = \eps(t) (\hatp^0\hatp^0)$.
Then
$$
K_0 \cdot \GI = \dot\eps(t) \psi_0^1 ,
$$
so according to (\ref{noetherK}) we have $r=\dot\eps(t)$,
and since $\gamma_\psi = \hatp^1$ we obtain
\beq
\delta \x = 2\eps(t) \hatp^0 + \dot\eps(t) \hatp^1
\eeq
---see
\cite{RR-planecurve}
for a geometric interpretaion of this transformation.

It can be shown that these transformations coincide with
those obtained in
\cite{RR-W3}
by considering a first-order lagrangian
when the supplementary variables are written in terms
of derivatives of~$\x$.

Notice that these transformations and their generating functions
$\GI$ are projectable to the hamiltonian space;
however, as we have explained at the end of the preceding section,
they do not yield hamiltonian gauge transformations,
as it can be easily checked.

\section{Other examples}

Here we consider two simple examples
of second-order singular lagrangians
to illustrate our procedure.

\paragraph{1.}
$L(x_0,x_1,x_2) = x_2$

The momenta are $\hat p^1 = 1$ and $\hat p^0 = 0$.

There are two hamiltonian constraints,
$\lligam21 = 1-p^1$ and $\lligam22 = p^0$.
In the intermediate space there is one constraint,
$\lligam11 = p^0$.
And finally there are no lagrangian constraints.

Let us look for a gauge Noether's transformation
``generated'' by a function $\GI = \eps(t) p^0$.
We obtain $K_0 \cdot \GI = 0$,
so it satisfies the required condition,
and the transformation is
$\delta x = \pot_0^*(\derpar{\GI}{p^0}) = \eps(t)$;
this says that $x(t)$ is completely arbitrary,
which of course is a consequence of the fact that $[L] = 0$
identically.

Notice that $\GI$ projectable to
a function $\GH = \dot\eps(p^1-1) + \eps p^0$ in the hamiltonian space.
For this function $K_1 \cdot \GH = 0$,
and so in this case we obtain a hamiltonian gauge transformation,
$\delta x^0 = \eps$, $\delta x^1 = \dot\eps$,
$\delta p^0 = \delta p^1 = 0$.

\paragraph{2.}
$\displaystyle L(x_0,x_1,x_2) = \frac12 (x_1 x_1)$

This is a first-order lagrangian,
but let us treat it as a second-order one.
The momenta are $\hat p^1 = 0$ and $\hat p^0 = x_1$.

In this example there are no lagrangian constraints.
In the intermediate space there is one constraint,
$\lligam11 = p^0$.
There are two hamiltonian constraints,
$\lligam21 = p^1$ and $\lligam22 = x_1-p^0$;
they are second-class.
In the intermediate space we have
$\lligam11 = x_1-p^0$ and $\lligam12 = x_2$.
And finally we obtain two lagrangian constraints,
$\lligam01 = x_2$ and $\lligam02 = x_3$.

As usual let us look for a function
$\GI = f x_2$.
Now we find that
$K_0 \cdot \GI = (K_0 \cdot f)x_2 + \pot_0^*(f) x_3$;
this is to vanish on the primary lagrangian constraint submanifold,
so necessarily we have $\pot_0^*(f) \feble{} 0$, $\GI \forta{} 0$
and therefore there are no Noether's gauge transformations;
this was expected since the solutions of the equations of motion
are paths of constant velocity.

Now let us look for the rigid Noether's transformations of this
lagrangian.
Due to the constraints of the intermediate space $P_1$,
we try a function $\GI(t,x_0,x_1)$.
We obtain
$$
K_0 \cdot \GI =
\derpar{\GI}{t} + x_1\derpar{\GI}{x_0} + x_2\derpar{\GI}{x_1} .
$$
Since this has to vanish on the surface $x_2 \feble{} 0$,
we obtain the condition
$\derpar{\GI}{t} + x_1\derpar{\GI}{x_0} = 0$,
from which
$\GI = g(x_1t-x_0,x_1)$;
this yields two independent transformations,
which are computed using the other term, the coefficient of $x_2$,
$r = \derpar{\GI}{x_1}$.

\section{Conclusions}

In this paper we have studied Noether's symmetries for higher-order
lagrangians.
This study is performed by using some intermediate spaces
between those of lagrangian and hamiltonian spaces.
We have seen that a conserved quantity of a Noether's transformation
can be characterized in terms of a function in the first intermediate
space satisfying a certain condition;
this is also useful to find gauge transformations when the lagrangian is
singular.

The issue of projectability to phase space of
the lagrangian conserved quantities as well as of
the transformations themselves
becomes quite more involved than in the first order case.  To get a
clearer picture of the subject we have made a thorough study of the
second-order case, where the structures of the general higher-order case
already show up.
As a consequence of this study, we present a variety of cases covering
all the possibilities with regard to
the projectability (or partial projectability) of the quantities
involved.

We give also some examples that illustrate several cases that appear in
our analysis. In particular, the example of section~5 does not possess
hamiltonian gauge generators, in spite of the fact that it has
lagrangian Noether's transformations which are projectable
to the hamiltonian space.

\paragraph{Acknowledgements}

We thank C. Batlle for bringing to our attention the possible
non existence of hamiltonian gauge generators for the model in
section~5.
We thank also J. Roca for useful comments about this model.

\appendix
\section{Gauge transformations in the hamiltonian formalism}

In this appendix we recall some results from
\cite{GP-gauge}.
We call {\it dynamical symmetry transformations}\/
those transformations which map solutions of
some equations of motion into solutions.

In the Dirac's hamiltonian formalism,
the {\it necessary and sufficient condition}\/
for a function $\GH(q,p;t)$ to generate,
through Poisson bracket,
$$
\delta f = \{f,\GH \} ,
$$
an infinitesimal dynamical symmetry transformation is
that $\GH$ be a {\it first class function}\/ and satisfy
\beasn
\{\GH  , H\} + \derpar{\GH }{t} & \forta{P^{(f)}} & \PFC
\\
\{\PFC, \GH \}                 & \forta{P^{(f)}} & \PFC ,
\label{dst}
\eeasn
where $P^{(f)}$ is the submanifold defined by all the hamiltonian
constraints in phase space,
\PFC\ stands for any primary first class hamiltonian constraint,
and the notation
$f \forta{M} 0$ means $f \feble{M} 0$ and $\dif f \feble{M} 0$
(Dirac's strong equality).

This conditions can be equivalently expressed in a more compact form:
$$
K \cdot \GH  \forta{V^{(f)}} 0 ,
$$
where $V^{(f)}$ is the surface
defined by all the lagrangian constraints in velocity space
and $K$ is the time-evolution operator $K$ for first-order lagrangians
---see for instance
\cite{GP-K}.
Though in
\cite{GP-gauge}
this is proved for first-order lagrangians,
it can be shown that this is also true for higher-order lagrangians.
More precisely, the condition is
\bea
K_{k-1} \cdot \GH \forta{\P{k-1}{f}} 0 ,
\label{kg}
\eea
where $\P{k-1}{f}$ is the surface
defined by all the constraints in the space~$P_{k-1}$.

\smallskip
More particularly, we call {\it gauge transformation}\/
a dynamical symmetry transformation
which depends on arbitrary functions of time.
The general form for a generator
of a hamiltonian gauge transformation,
depending on one arbitrary function,
can be taken as
\beq
\GH(q,p;t) = \sum_{k \geq 0} \epsilon^{(-k)}(t) \, G_k(q,p) ,
\label{G}
\eeq
where $\epsilon^{(-k)}$ is a $k$th primitive
of an arbitrary function of time~$\epsilon$.

To find a gauge generator,
the characterization (\ref{kg}) or (\ref{dst})
of $\GH$ as a dynamical symmetry generator
splits yielding the following constructive algorithm,
where strong equalities have been changed to normal equalities
\cite{GP-gauge}:
\beasn
G_0                  & =         & \PFC
\\
\{G_k, H\} + G_{k+1} & =         & \PFC
\\
\{\PFC, G_k\}        &\forta{P^{(f)}}& \PFC .
\label{algor}
\eeasn
It is noticed, therefore,
that though there may be second class constraints,
the generators of hamiltonian gauge transformations
are built up of {\it first class constraints},
and, according to (\ref{algor}.a),
are headed by a primary one.

Some results on the existence of a basis of
primary first class hamiltonian constraints
each one yielding a gauge transformations are known:
this is guaranteed
under some regularity conditions
\cite{GHP-exist},
namely
the constancy of the rank of Poisson brackets among constraints
and the non appearance of ineffective constraints.
If these hamiltonian gauge transformations exist,
their pull-back constitutes
a complete set of lagrangian gauge transformations.

On the other hand,
as we have said in the introduction,
there are examples of first-order lagrangians for which
hamiltonian gauge generators do not exist,
whereas they have lagrangian gauge transformations~%
\cite{GP-Noether}.
In this paper we have seen that this also happens for a
relativistic particle with lagrangian proportional to the curvature.


\end{document}